\documentstyle[12pt]{article}
\newcommand{\ba}{\begin{array}}
\newcommand{\ea}{\end{array}}

\newcommand{\be}{\begin{equation}}
\newcommand{\ee}{\end{equation}}
\newcommand{\bea}{\begin{eqnarray}}
\newcommand{\eea}{\end{eqnarray}}
\newcommand{\beal}{\setcounter{letter}{1} \begin{eqnarray}}
\newcommand{\eeal}{\addtocounter{equation}{1} \end{eqnarray}}
\newcommand{\none}{\nonumber \\}
\newcommand{\req}[1]{Eq.(\ref{#1})}

\newcommand{\larrow}{\,\,\,\,\hbox to 30pt{\rightarrowfill}
\,\,\,\,}
\newcommand{\slarrow}{\,\,\,\hbox to 20pt{\rightarrowfill}
\,\,\,}

\newcommand{\half}{{1\over2}}

\begin{document}
\mbox{}\\[1cm]
\begin{center}
{\bf
OBSERVABLES FOR TWO-DIMENSIONAL BLACK HOLES}
\end{center}
\vspace{0.5 cm}
\begin{center}
{\sl by\\
}
\vspace*{0.50cm}
{\bf
J. Gegenberg $\dagger$
G. Kunstatter $\sharp$ and D. Louis-Martinez $\sharp$
\\}
\vspace*{0.50cm}
{\sl
$\dagger$ Dept. of Mathematics and Statistics,
University of New Brunswick\\
Fredericton, New Brunswick\\
Canada  E3B 5A3\\
{[e-mail: lenin@math.unb.ca]}\\ [5pt]
}
{\sl
$\sharp$ Dept. of Physics and Winnipeg Institute of
Theoretical Physics, University of Winnipeg\\
Winnipeg, Manitoba\\
Canada R3B 2E9\\
{[e-mail: uowgxk@ccm.umanitoba.ca]}
}
\end{center}
\bigskip\noindent
{\large
ABSTRACT
}
\par
\noindent
We consider the most general dilaton gravity theory in 1+1
dimensions. By suitably parametrizing the metric and scalar field
we find a simple expression that relates the energy of a generic
solution to the magnitude of the corresponding Killing vector. In
theories that admit black hole solutions, this relationship leads
directly to an expression for the  entropy $S=2\pi \tau_0/G$,
where $\tau_0$ is the value of the scalar field (in this
parametrization) at the event horizon. This result agrees with the
one obtained using the more general method of Wald. Finally, we
point out an intriguing connection between the black hole entropy
and the imaginary part of the ``phase" of the exact Dirac quantum
wave functionals for the theory.\\[10pt]
gr-qc/9408015\\
WIN-94-07
\newpage
\renewcommand{\baselinestretch}{1.5}
\section{Introduction}\medskip
Two dimensional theories of gravity have been the subject of much
interest for a number of years because of their connection to
string theory and their interesting mathematical properties. There
has been an explosion of work in this area in the last few years
due to the discovery by Callan {\it et. al.} \cite{CGHS} that such
stringy theories may
provide models
for black hole evaporation in which fundamental questions
concerning
the endpoint of collapse could in principle be addressed
rigorously, if not exactly.
\par
Most theories that have been considered contain one scalar field
 plus the graviton field. The most such general coordinate
invariant theory
(with at most two derivatives) has been examined by Banks and
O'Laughlin\cite{banks} and subsequently by others\cite{mann}.
Special
cases of current interest include the
string-derived model, the Jackiw-Teitelboim model\cite{jackiw}
and spherically symmetric gravity\cite{spherical}.
\par
Here we consider the classical observables in the most general 2-D
dilaton gravity theory.  Our main interest
is in theories that have black hole solutions. By choosing a
convenient parametrization for the scalar field and metric tensor,
it is possible to write a very simple coordinate
invariant expression for the Killing vector in the general theory.
This can then be used to shed considerable light on the remaining
observables in the theory. For example we prove that the coordinate
invariant constant parametrizing the solutions is the conserved
quantity associated with translations
along the Killing direction (i.e. the energy).
We also show that the momentum conjugate to the energy is
the (Killing)-time separation at infinity. This has been shown by
Kuchar\cite{kuchar94} for spherically symmetric gravity. In
addition, knowledge of the Killing vector enables us to calculate
the
surface gravity for a generic 2-D black hole, and derive
a simple expression relating the energy to the value of the scalar
field at the horizon. This leads to a very simple
derivation of the entropy for a generic 2-D black hole. From this
expression we are able to show a deep connection between the
entropy, and the imaginary part of the phase of the physical
quantum wave functional, which was derived for the general theory
in \cite{domingo1}.
\section{\bf Action and Killing Vector}\medskip
The most general action functional depending on the metric tensor
$g_{\mu\nu}$ and a scalar field $\phi$ in two spacetime dimensions,
such that it contains at most second derivatives of the fields can
be written\cite{banks}:
\be
S[g,\phi]={1\over2G}\int d^2x \sqrt{-g} \left( \half
g^{\alpha\beta}
\partial_\alpha \phi \partial_\beta \phi - {1\over
l^2}\tilde{V}(\phi) +
D(\phi)
R\right).
\label{eq: action 1}
\ee
The metric, scalar field and 1+1 dimensional gravitation constant
$G$ are assumed to be dimensionless. This requires the introduction
of a dimensionfull parameter into the potential. We have chosen to
make this parameter explicit in the Lagrangian, since it plays an
important role in determining the dimensionally correct physical
observables in the generic theory. For spherically symmetric
gravity, $l= l_p$ is the Planck length.
As first discussed in \cite{banks} and shown explicitly in
\cite{domingo1}, by reparametrizing the  fields:
\bea
 g_{\mu\nu} &\to& h_{\mu\nu}=\Omega^2(\phi) g_{\mu\nu},\\
\phi &\to& \tau=D(\phi),
\label{eq: gbar defn}
\eea
with
$\Omega^2 = \exp \half\int {d\phi \over (dD/d\phi)}$ one can
eliminate the kinetic
term for the scalar and put the action in the form:
\be
I[h,\tau]={1\over2G}\int_{M^2}d^2x \sqrt{-h}\left(\tau R(h)-
{1\over l^2}{V(\tau)}
\right).\label{eq: dilaton action}
\ee
where $V(\tau)$ is an arbitrary function of the scalar field
$\tau$.
\par
The equations of motion take the simple form:
\be
R={1\over l^2} {dV\over d\tau} \label{eq:eofm1}
,\ee
and
\be
\nabla_\mu\nabla_\nu\tau+{1\over 2l^2} g_{\mu\nu}V=0
\label{eq:eofm2}.
\ee
The most general solution to these equations has been
found\cite{domingo2}. In the convenient gauge:
\be
\tau=x/l, g_{tx}=0,
\ee
the solution is:
\be
ds^2=-(-{J(x/l) }-C)dt^2+(-{J(x/l)}-C)^{-1}dx^2 \,\,\, ,
\label{eq: generic solution}
\ee
where $J'(\tau)= V(\tau)$ and $C$ is a
coordinate-invariant constant of integration that characterizes
the
physically distinct solutions in the theory. It can be expressed
in covariant form:
\be
C = -|\nabla\tau|^2 l^2- J(\tau) \,\, .
\label{eq: covariant C}
\ee
We will show later that $C/2l$ is the energy of the solution.
\par
Since the solutions given above depend only on the spatial
coordinate, clearly they each have a Killing vector, so that the
generalization to Birkhoff's theorem holds for 2-D dilaton gravity,
as shown in \cite{domingo2}. The Killing vector can in fact be
written in any coordinate system as:
\be
k^\mu=l\eta^{\mu\nu}\tau,{}_\nu \label{eq:killing} \,\, .
\label{eq: killing vector}
\ee
In the above
$\eta^{\mu\nu}=-\eta^{\nu\mu}={1\over\sqrt{-g}}\epsilon^{\mu\nu}$
is the antisymmetric {\it tensor}. The constant $l$ is required to
make the Killing vector components dimensionless.
It can easily be verified that \req{eq:eofm2} implies that $k^\mu$
satisfies the Killing equation $\nabla_{(\mu}k_{\nu)}=0 $ on shell.
Moreover, it is clear that
$\tau_{,\mu}k^\mu=0$ identically, so that the scalar field is also
invariant along the Killing directions. Note that
\be
\mid k\mid^2=-l^2\mid\nabla\tau\mid^2\none
            = C + J(\tau)  \,\, .
\label{eq:magkilling}
\ee
\par
\par
The question of which of the generic dilaton
theories
admit black hole solutions can be addressed at this
point\cite{mann}.
A necessary condition that a model admit a black hole configuration
is that there exist at least one curve in spacetime given by
$\tau(x,t)=\tau_0=constant$, such that $J(\tau_0)=-C$. In addition,
$J(\tau)$ must be monotonic (in $\tau$) in a neighbourhood of
$\tau_0$.
\par
Before closing this section, we will display the various quantities
defined here in the special case of spherically symmetric gravity,
for which $V(\tau)=-1/\sqrt{2\tau}$.  The static solution for the
metric in our parametrization  is related
to the usual Schwarzschild solution by the conformal
reparametrization $ds^2 = \sqrt{2\tau}ds^2_{schwarz}$. In terms of
the coordinate
$r=l\sqrt{2\tau}$, the metric \req{eq: generic solution} takes the
form:
\be
h_{\mu\nu}dx^\mu dx^\nu={r\over l}\left\{-(1-2m/r)dt^2+(1-2m/r)^{-
1}\right\}dr^2,
\ee
where the mass $m=l C/2$. Finally, $(k^\mu)=(1,0)$ and
$|k|^2 = (2m-r)/l$.
\bigskip\noindent
\section{Hamiltonian Analysis}
\medskip
We now review a Hamiltonian analysis of the general
1+1--dimensional theory\cite{domingo1}.
Spacetime is split into
a product of space and time:  $M_2 \simeq
\Sigma\times
R$ and the metric $h_{\mu\nu}$ is given an ADM-like
parameterization:\cite{torre}
\be
ds^2=e^\alpha\left[-\left(M^2+N^2\right)dt^2+\left(dx+Mdt\right)
^2\right]
.\label{eq:adm}
\ee
where $\alpha$, $M$
and
$N$ are functions on spacetime $M_2$.
We define the quantity $\sigma$ by $\sigma^2:= M^2+N^2$.  Also, in
the following, we denote by the overdot and prime, respectively,
derivatives with respect to the time coordinate $t$ and spatial
coordinate $x$.
\par
The canonical momenta for the fields
$\{\alpha,\tau\}$ are respectively:
\bea
\Pi_\alpha&=&{1\over2G\sigma}\left(M\tau'-\dot\tau\right),
\label{eq: pi_alpha}
\\
\Pi_\tau&=&{1\over2G\sigma}\left(-\dot\alpha+M\alpha'+2M'
\right),
\label{eq: pi_tau}
\eea
The vanishing
of the momenta canonically conjugate to $M$ and $\sigma$
yield the primary constraints for the system. Following the
standard
Dirac prescription\cite{dirac}, we obtain the canonical
Hamiltonian (up to spatial divergences):
\be
H_0=\int dx\left(M{\cal F}+{1\over2G}\sigma{\cal
G}\right).\label{eq: ham}
\ee
where we have defined:
\be {\cal F}:=\alpha' \Pi_\alpha+\tau' \Pi_\tau-2\Pi_\alpha'
\label{eq: gauss} \ee
\be
{\cal
G}:=2\tau''-\alpha'\tau'-\left(2G\right)^2
\Pi_\alpha
\Pi_\tau+{1\over l^2}{e^\alpha}V(\tau).\label{eq: hamcon}
\ee
Clearly ${1\over2G}\sigma$ and $M$ play the role
of Lagrange multipliers that enforce the secondary constraints
${\cal F}\approx 0$ and ${\cal G}\approx 0$.
\par
The energy can be constructed by
noting that the
following linear combination of the constraints is a total spatial
derivative:
\bea
\tilde{\cal G}&:=&{l\over 2} e^{-\alpha}\left((2G)^2\Pi_\alpha{\cal
F}+
\tau'{\cal G}\right)\none
&=&(q[\alpha,\tau,\Pi_\alpha,\Pi_\tau])'
\approx 0 \,\, ,
\label{eq: def'n q'}
\eea
where we have defined the variable $q$ as
\be
q:={l\over 2}\left[e^{-\alpha}\left((2G\Pi_\alpha)^2-
(\tau')^2\right)- l^{-2}J(\tau)\right].\label{eq:qdef}
\ee
The expression on the right-hand side above is nominally an
implicit function
of the spatial
coordinate, but is constant on the constraint surface. Moreover,
it is straightforward to show
that $q$ commutes with both
constraints ${\cal F, G}$. Thus, the constant mode of $q$ is a
physical
observable in the Dirac sense.
\par
In terms of the canonical momenta the magnitude of the Killing
vector can be written as:
\be \mid
k\mid^2=l^2e^{-\alpha}\left[(2G\Pi_\alpha)^2-(\tau')^2\right] \,\,
{}.
\label{eq:kill/mom} \ee
Thus the observable $q$ is:
\bea
q&=&{1\over 2l}\left(\mid k\mid^2-J(\tau)\right)\none
  &=& {C\over 2l} \,\, .
\label{eq:cov q}
\eea
It is worth noting that the constancy of $q$ in {\it spacetime}
follows
by contracting the field equations \req{eq:eofm2} by
$\nabla^\mu\tau$.
\par
We now prove that the generator $\tilde {\cal G}= q'$
generates diffeomorphisms along the direction of the Killing vector
$k^\mu$.   We consider an infinitesimal translation
$x^\mu\to x^\mu+f^\mu$, where $f^\lambda:=-vk^\lambda$.  Using the
ADM
parameterization
\req{eq:adm}, we find on the constraint surface that
\bea
\delta\tau&=&0,\\
\delta\alpha&=&4Gle^{-\alpha}\Pi_\alpha v'.
\eea
On the other hand, the transformations generated by
$\tilde{\cal G}(v)$ are:
\bea
\bar\delta\tau&:=&\{\tilde{\cal G}(v),\tau(x)\}= 0,\\
\bar\delta\alpha&=&\{\tilde{\cal G}(v),\alpha(x)\}\\
&=& (2G)^2 l e^{-\alpha}\Pi_\alpha v'.
\eea
Comparing these two transformations one finds that
the observable $q/G$ is the conserved
quantity associated with translations along the Killing vector;
i.e. it is the energy. This can also be verified by writing the
canonical Hamiltonian in terms of $\tilde{\cal G}$. One finds:
\be
H_0 = -\left({\dot{\tau}\over \tau'}\right) {\cal F}
   + \left( {\sigma e^\alpha \over l \tau'}\right){q'\over G} \,\,
{}.
\ee
In order to obtain Hamilton's equations, it is
necessary to add the following surface term to the canonical
Hamiltonian:
\be
H_{ADM} = \int dx \left(\left( {\sigma e^\alpha \over l
\tau'}\right){q\over G}\right)' \,\, .
\ee
It is easy to verify that for solutions of the form \req{eq:
generic solution},
${\sigma e^\alpha / l \tau'}=1$. Hence, $H_{ADM} = q/G$ is the ADM
energy,
as expected.
\par
The momentum conjugate to $q$, is found by inspection to
be\cite{domingo1}:
\be
p:=-\int_\Sigma dx {2\Pi_\alpha e^\alpha\over
(2G\Pi_\alpha)^2-(\tau')^2} \,\, .
\label{eq:pdef}
\ee
It can easily be verified that the Poisson algebra for the fields
and the momenta leads directly to $\{q,p\} = 1$. Under  a
general gauge transformation \bea
\delta p &:=& \{{\cal G}(v)+{\cal F}(w),p\}\none
&=&-\int dx \left( \frac{e^{\alpha}(w \Pi_\alpha -v
\tau')}
{(2G\Pi_\alpha)^2 -  (\tau')^2} \right)' \,\,.
\label{j}
\eea
Thus $p$ is gauge invariant only if the test functions $v$ and $w$
vanish sufficiently rapidly at
infinity. The value of $p$ depends on the global properties of the
spacetime slicing. This is consistent with the generalized Birkhoff
theorem\cite{domingo1} which states that there is only one
independent diffeomorphism invariant parameter
characterizing the space of solutions.
\par
It is instructive to write the observable $p$ in covariant form:
\bea
p&=&\int_\Sigma dx e^{\alpha/2}n^\mu{\nabla_\mu\tau\over\mid
k\mid^2}\\
  &=&-2\int_\Sigma dx^\mu {k_\mu\over\mid
k\mid^2}.\label{eq: cov p}
\eea
Note that $dxe^{\alpha/2}$ is the measure induced on $\Sigma$ by
$h_{\mu\nu}$.
In the expression for $p$ the vector field $n^\mu$ is the unit
(timelike)
normal to $\Sigma$.
\par
Using \req{eq: cov p} it is straightforward to show that  the
global variable $p$ is the time separation at infinity of
neighbouring spacelike
surfaces which are asymptotically normal to the Killing vector
field $k^\mu$.  We suppose that $V(\tau)$ is such that in the
region exterior to the event horizon, $k^\mu$ is
timelike.  Let $U$ be the ``triangular region" of
spacetime bounded by spacelike surfaces $\Sigma_1, \Sigma_2$ and
by a timelike surface $T$ at infinity tangent to $k^\mu$.  It is
straightforward to show that
$\nabla_\mu\left(\nabla^\mu\tau/\mid
 k\mid^2\right)\equiv0$ in $U$ for any solution of the equations
of motion.  Hence by Gauss' Theorem
 \bea
0&=&\int_Ud^2x\nabla_\mu\left({\nabla^\mu\tau\over\mid
k\mid^2}\right)\none
&=&p_2-p_1 +\int_T d\mu[T]t^\mu {\nabla_\mu\tau\over \mid k\mid^2},
\eea
where $p_1, p_2$ are the values of $p$ on $\Sigma_1, \Sigma_2$,
respectively, $d\mu[T]$ is the measure on $T$ and $t^\mu$ is the
outward unit normal to $T$.  Now at infinity, the integral over $T$
above is just the time-separation of the spacelike surfaces.
Indeed, by definition, $t^\mu=\tau^{,\mu}/|k|^2$.  Choose the
measure
$d\mu[T]=\sqrt{h_{\theta\theta}}d\theta$, where $\theta$ is the
parametrization of the timelike line $T$ such that the induced
metric $h_{\theta\theta}:=h_{\mu\nu}{\partial
x^\mu\over\partial\theta}{\partial x^\nu\over\partial\theta}=\mid
k\mid^2$.  From this we immediately get the desired result.
\par
Finally, we observe that the integrand of the
observable $p$ has a pole at the location of any event horizon in
the model. Thus, analytic continuation is in general required to
make the expression well defined, and may introduce an imaginary
part to the observable $p$. For example, in spherically symmetric
gravity, one can show that in Kruskal coordinates the observable
$p$ integrated along a slice of constant Kruskal time $T$ takes the
simple form:
\bea
p&=&{2m\over G}\int dX\left[{1\over X-T} - {1\over X+T}
\right]\none
 &=&{2m\over G} \left.\ln\left({X-T\over X+T}\right)
\right|^{X_f}_{X_i} \,\,.
\label{eq: p in Kruskal}
\eea
$p$ is therefore precisely the difference in Schwarzschild times
at the
endpoints of the spatial slice. In this case there are simple poles
at $X=\pm T$ (i.e. at $r=2m$), so that for an eternal black hole,
with
suitable analytic continuation, Im${p}= 2\pi m/G$. Although this
potential imaginary piece is irrelevant classically for the
Schwarzschild time, it may have some significance in the quantum
theory in which $p$ is a physical phase space observable. This will
be discussed below.
\bigskip
\section{Thermodynamical Properties}
\medskip
We now calculate the surface gravity and entropy of a generic 2-D
black hole. The surface gravity $\kappa$
is determined by the following expression,
evaluated at the event horizon\cite{wald2}:
\be
\kappa^2=-{1\over2}\nabla^\mu k^\nu\nabla_\mu k_\nu.
\ee
Using \req{eq: killing vector} for $k^\mu$ and the field equations
\req{eq:eofm2} it is straightforward to show that:
\be
\kappa = - {1\over2 l}V(\tau_0) \,\, ,
\label{eq: kappa}
\ee
where $V(\tau_0)$ is the potential evaluated at $\tau=\tau_0$ (i.e.
on the event horizon).
The sign in \req{eq: kappa} was chosen to yield a positive surface
gravity for positive energy. Note that $\tau_0$ is given implicitly
as a function of the energy $q$ by requiring $|k|^2=0$ in
\req{eq:cov q}.
\par
The Hawking temperature for the generic black hole solution can
easily be calculated by defining the Euclidean time $t_E = it$ in
\req{eq: generic solution} and then finding the periodicity
condition on $t_E$ that makes the solution everywhere regular. This
is done by defining the coordinate $R^2:= -a (J(\tau) +C)$  and
choosing the constant $a$ so that the spatial part of the metric
goes to $dR^2$ at the event horizon $\tau_0$. A straightforward
calculation gives $a=|2l/V(\tau_0)|$, so that the Hawking
temperature,
which is the inverse of the period of $t_E$, is:
\be
T_H = {1\over 2\pi} {V(\tau_0)\over 2l} = {\kappa\over 2\pi} \,\,
,
\ee
as expected. Note that this calculation does not depend on the
details of the model: it merely requires the existence of a horizon
at which $J(\tau_0) = -C$
\par
The entropy, $S$, can now easily be determined by inspection of
\req{eq:cov q}. In particular, if we vary the solution, but stay
on the event horizon, we find that the variation of the energy is:
\be
\delta {\cal E}= \delta (q/G) = -{1\over 2l G} V(\tau_0) \delta
\tau_0 \,\, .
\label{eq: delta q}
\ee
Identifying the Hawking temperature and surface gravity derived
above, we find that the first law of thermodynamics
$\delta E = T\delta S$ will be satisfied providing we identify
the entropy to be
\be
S = {2\pi\over G} \tau_0 \,\, .
\ee
\par
Recently Wald \cite{wald} formulated a local geometric expression
for the
entropy of a black hole in any Lagrangian-based theory which admits
black hole
solutions.  Following is a brief summary of this construction.
\par
Denote any dynamical fields in the theory by $\phi$.
 Under a
diffeomorphism
generated by $v^\mu$, the Lagrangian, $L$, considered as a two
form, transforms as
\be
\delta L=E\cdot {\cal L}_v\phi + d\Theta,
\ee
where the product in the first term includes a summation over the
dynamical
fields and contraction over the tensor indices.  The components
of $E$
are just the Euler-Lagrange expressions for the action, and hence
the first
term vanishes on-shell.  The second term is the exterior derivative
of
a one-form field $\Theta$, which depends on $\phi$ and ${\cal
L}_v\phi$. From the identity ${\cal L}_v\gamma=v\cdot d\gamma
+d(v\cdot
\gamma)$ for
any differential form $\gamma$, the invariance of the action under
diffeomorphisms implies that on-shell the expression
\be
j:=\Theta-v\cdot L,
\ee
is closed.  Furthermore it can be demonstrated \cite{lee} that
on-shell
$j$ is exact, i.e. $j=dQ$, where $Q$ is a 0-form, locally
constructed from
the dynamical fields and their Lie derivatives with respect to $v$.
\par
If black hole solutions exist, Wald showed that the quantity
\be
S:={2\pi\over\kappa}{\cal Q}(x_0),
\ee
behaves like the entropy of the black hole.
For the generic dilaton gravity models, it can be shown that:
\bea
{\cal Q}&=&{1\over 2G}\eta_{\mu\nu}\left(2 v^\mu\nabla^\nu \tau +
\tau\nabla^\mu v^\nu\right) \,\, ,
\eea
where $v$ is an arbitrary
diffeomorphism.
Now for the case that $v^\mu=k^\mu=l\eta^{\mu\nu}\nabla_\nu\tau$,
it
follows that
the Wald's expression for the entropy is
\be
S={2\pi\over G}\tau(x_0) \,\, ,
\ee
in agreement  with the result obtained above. It also gives the
correct answer for spherically
symmetrical gravity (for which $\tau_0= 2 m^2/l^2_p$) and agrees
with the results obtained by
Frolov
\cite{frolov} and Iyer and Wald
\cite{iyer} for string motivated models.
\par
It is perhaps worth noting that the very simple expressions given
above for the Killing vector, surface gravity and entropy are only
valid in the given parametrization, which was obtained from the
generic form by a conformal reparametrization of the metric.
It is therefore worthwhile to ask how such conformal
reparametrizations affect the physical quantities described above.
First of all, the Killing vector is invariant under such a
transformation, since for solutions, the conformal factor
$\Omega^2(\tau)$ is also invariant along the Killing directions.
A straightforward calculation shows that the surface gravity for
a given solution is also unchanged. Since the energy, $q$,  which
is the conserved quantity associated with translations along the
Killing direction is also presumably invariant under
reparametrizations of the
fields (that leave the Killing vector invariant), the above
arguments would lead to precisely the same value
for the  entropy in any parametrization (although the dependence
of the entropy on the fields will in general be considerably more
complicated in different parametrizations).
\par
The exact quantum wavefunctional which solves the constraints with
a particular factor ordering has been found to be\cite{domingo1}:
\be
\psi_{phys}[q;\alpha,\tau]= exp{i\over G} \chi[q;\alpha,\tau] .
\ee
where the ``phase" is given by:
\be
\chi[q;\alpha,\tau] = \int dx \left[Q+ {\tau'\over2} \ln
\left({\tau' -
    Q\over \tau'+Q}\right)\right] \,\, ,
\label{eq: wave function}
\ee
with
\be
Q:= \sqrt{ (\tau')^2+ e^\alpha\left({2q\over l}+{J(\tau)\over
l^2}\right)}
\ee
which is equal to $(2G\Pi_\alpha)$ on the constraint surface.
If we
restrict to classically allowed regions, for which $Q^2\ge 0$ then
the phase $S$ can acquire an imaginary part from the logarithm when
$(\tau')^2 - Q^2\le 0$. This is precisely the region where the
Killing vector for the solution is spacelike (in a non-singular
coordinate system for which $e^\alpha$ is positive).
Therefore for theories with an event horizon, the logarithm in
\req{eq: wave function} can be analytically continued so that
\be
Im \chi = {i\pi   \tau_0\over 2} = i {S\over 4} \,\, .
\ee
The
imaginary part of the ``phase" is therefore proportional to the
entropy of the black hole. This is consistent with an earlier
heuristic result obtained for spherically symmetric
gravity\cite{gk2}.
\section{Conclusions}
\medskip
 We have shown that in a suitable
parametrization, the Killing vector for a generic 2-D black hole
takes a particularly simply
form, and can be used to shed considerable light on the nature of
the physical observables in the theory. In particular we were able
to show that the space of physical observables consists of two
conjugate variables: the energy and the Killing time-separation at
infinity. The former is the conserved quantity associated with
translations along the Killing direction. The latter depends
explicitly on the global properties of the space-time slicing, as
required by the generalized Birkhoff's theorem valid for such
theories.  Moreover, in the calculation of the time-separation, it
is necessary to analytically continue through the event horizon (in
those models which display this feature). We also used the explicit
expression for the Killing vector to calculate the surface gravity
and entropy for the general theory. The latter agrees with the
result obtained using Wald's more general method. Finally we
showed an intriguing relationship between the entropy and imaginary
part of the phase of the exact quantum wave functional in the Dirac
quantized theory.
function).
\par
We therefore believe that the above formalism provides a powerful
tool for analyzing the classical, quantum and  thermo- dynamics of
generic 1+1 dimensional black holes.
 \par\vspace*{10pt}
\noindent
{\large\bf Acknowledgements}
\par
The authors are grateful to
A. Barvinsky, S. Carlip, V. Frolov, D. McManus, P. Sutton, G.A.
Vilkovisky and D.  Vincent for helpful discussions.  This work
was supported in part by the Natural Sciences and Engineering
Research
Council of Canada.  \par
\end{document}